\documentclass[aps,prl,twocolumn,floats,10pt]{revtex4}
\usepackage{graphicx}
\usepackage{amssymb}
\usepackage{amsmath}
\usepackage{float}
\usepackage{color}

\begin{document}

\title{Interaction of Electromagnetic Radiation with Luminal Mirror}

\author{T. Z. Esirkepov$^a$ and S. V. Bulanov$^{a,b}$}
\affiliation{$^a$Kansai Institute for Photon Science, National institutes for Quantum and Radiological Science and Technology (QST),
8-1-7 Umemidai, Kizugawa, Kyoto 619-0215, Japan\\
$^b$Extreme Light Infrastructure ERIC, ELI--Beamlines Facility, Za Radnici 835, Dolni Brezany 25241, Czech Republic
}

\begin{abstract}
A modulation of refractive index can move at the speed of light.
How it interacts with an electromagnetic wave? Does it reflect?
We show that an incident electromagnetic wave, depending on its frequency
either is totally transmitted with a phase shift,
or forms a standing wave, or is totally reflected with the frequency upshift.
A short incident pulse is converted into a wavepacket that has all three parts
(transmitted, standing and reflected waves).
The reflected part near the interface exhibits an infinitely growing in time local frequency.
The wavepacket's energy spectral density asymptotically is the inverse square of frequency.
If the refractive index modulation disappears, the high frequency radiation is released.
\end{abstract}

\maketitle

An electromagnetic wave reflected off a moving mirror
acquires modified frequency, $\omega_r$, and electric filed, $E_r$,
as predicted by A. Einstein \cite{Einstein1905}:
\begin{equation}\label{eq-Einstein}
\omega_r/\omega_i=E_r/(r E_i) = (1+\beta)/(1-\beta).
\end{equation}
Here $\omega_i$ and $E_i$ are, respectively, the incident wave frequency and electric field strength;
$r$ and $\beta$ are the mirror reflectivety and velocity normalized to the speed of light in vacuum.
The reflected radiation either gets energy from the approaching mirror ($0<\beta<1$),
or transfers energy to the receding mirror ($-1<\beta<0$).
Both cases have important applications.
The former is essential in obtaining high-frequency and ultra-intense radiation
\cite{High-freq}.
The latter underlies the radiation pressure dominant acceleration (RPDA) of ions
\cite{RPDA}.

When the mirror velocity gets closer to the speed of light in vacuum, the reflected wave frequency formally tends to infinity. On the one hand, the mirror reflectivity vanishes, because in its proper reference frame the incident electromagnetic wave frequency also tends to infinity rendering any matter transparent. On the other hand, the refractive index modulation inside a thick mirror can be independent of the mirror-vacuum interface position, at least during some finite time interval.
This leads to a paradox: in the limit $\beta\rightarrow 1$ a thick mirror is totally transparent merely because of its interface motion, but the bulk of the mirror may not always depend on that motion and can refract.
To resolve this paradox, we analyze the electromagnetic wave propagation in medium where the refractive index modulation extends at the speed of light in vacuum.

Luminal or superluminal velocity \cite{Superlum} of an object emitting or refracting electromagnetic radiation
does not contradict to the special theory of relativity,
as long as this is not a massive particle velocity or a wave group velocity \cite{BG}.
For example, ionization induced in gases by an external electromagnetic beam rotating like a searchlight can produce a spot of a non-zero Langmuir frequency moving with an arbitrary velocity \cite{Ioniz}.
Accordingly, the terms ``luminal plasma-vacuum interface or plasma slab'' used below
do not imply a luminal motion of plasma particles. 

{\bf General solution.}
We start from the Maxwell equation for the electromagnetic waves propagating in inhomogeneous plasma:
\begin{equation}\label{eq-Axt}
\partial_{tt}A-c^2 \partial_{xx}A+\omega_{pe}^2 \rho(x/c-\beta t) A =0.
\end{equation}
Here $A(t,x)$ is the vector-potential’s transverse component;
$\omega_{pe}={\rm const}$ is the Langmuir frequency for $\rho=1$;
$\rho$ is the dimensionless profile function;
$c$ is the speed of light in vacuum;
$\beta>0$ is the velocity normalized to $c$.
We assume that $\left|A(t,x)\right|$ is sufficiently small,
so that we neglect a longitudinal electric current generation,
including the recoil effects and magnetic field induction.
In this approximation, the electric field strength is
$E=-\partial A/\partial t$.

\begin{figure}
\includegraphics[width=0.8\columnwidth]{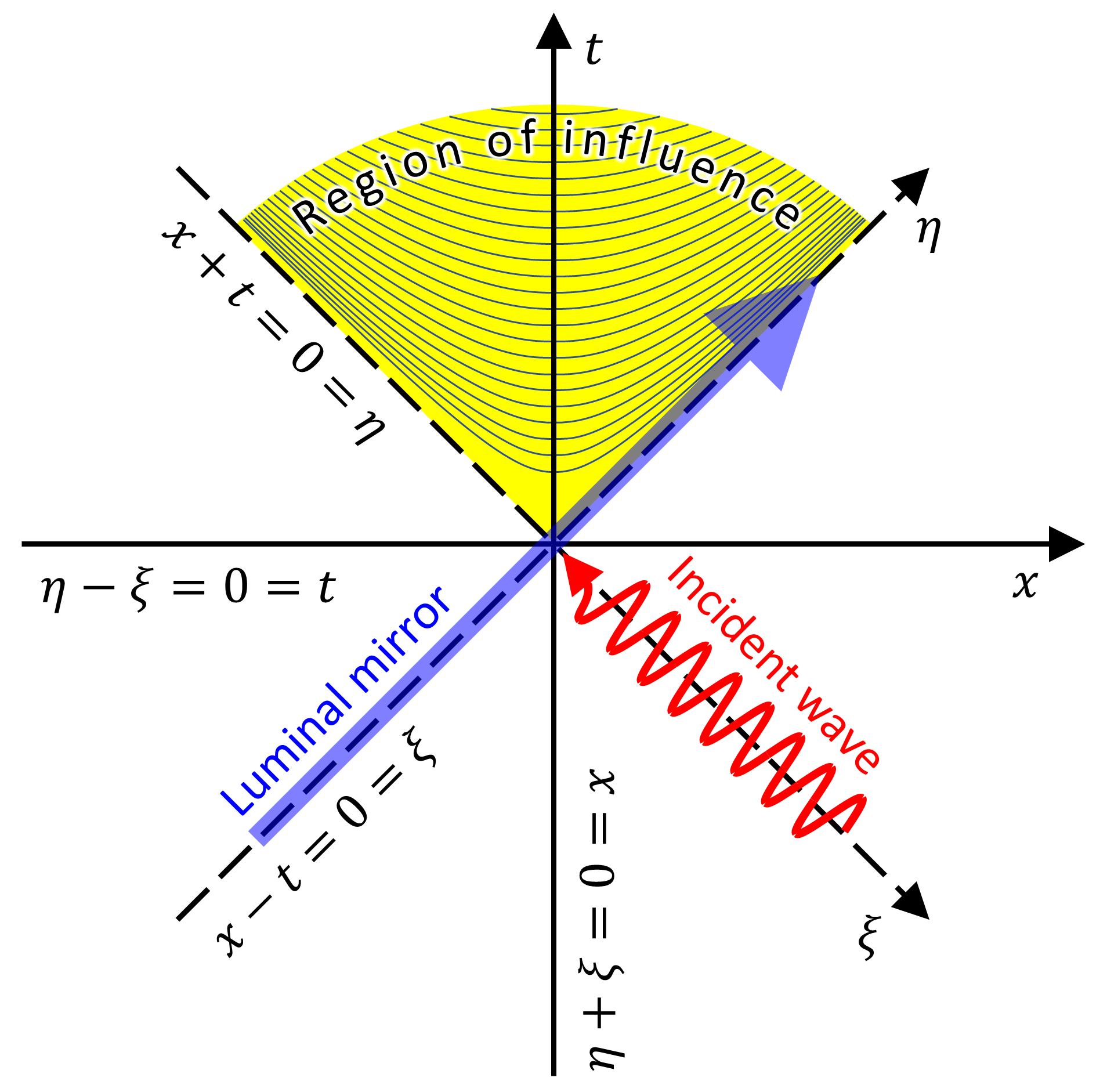}
\caption{%
Variables $(x,t)$ and $(\zeta,\eta)$,
and the region of influence of Eq. \ref{eq-Axt} for $\beta=1$.%
\label{fig-0}
}
\end{figure}

Changing to the new variables $\bar\zeta=x/c -\beta t, \eta=x/c + t$,
and performing the Fourier transform with respect to $\eta$ according to
$a(\bar\zeta) = \frac{1}{\sqrt{2\pi}}\int_{-\infty}^{+\infty}A(\bar\zeta,\eta) e^{i\Omega\eta}d\eta$,
we reduce Eq. (\ref{eq-Axt}) to the ordinary differential equation for $a(\bar\zeta)$
with $\Omega$ as a parameter:
\begin{equation}\label{eq-a-xi}
(\beta^2-1)a'' + 2i\Omega(1+\beta)a' + \omega_{pe}^2 \rho(\bar\zeta)a = 0.
\end{equation}
%
%
According to the Tikhonov theorem \cite{Tikhonov},
in the limit of $\beta\rightarrow 1$ the solution of Eq. (\ref{eq-a-xi}) tends to the
solution of the degenerate equation
\begin{equation}\label{eq-a1}
4i\Omega a'(\zeta) + \omega_{pe}^2 \rho(\zeta)a(\zeta) = 0, \;\;\; \zeta=x/c-t,
\end{equation}
that is $a(\zeta) = a_0(\Omega) e^{i\xi/\Omega}$
with $\xi=(\omega_{pe}/2)^2 S(\zeta)$ and $S'(\zeta)=\rho(\zeta)$.
The initial condition determines $a_0(\Omega)$ and the integration constant in $S$.
The inverse Fourier transform of $a(\zeta)$ with respect to $\Omega$
gives the solution 
\begin{equation}\label{eq-Asol}
A(\xi,\eta) = \frac{1}{\sqrt{2\pi}}
\int_{-\infty}^{+\infty}\!\!\!\! a_0(\Omega)\exp\left[i\left( \frac{\xi}{\Omega} - \Omega\eta \right)\right]d\Omega.
\end{equation}
Direct differentiation with respect to $\xi$ and $\eta$ 
confirms that $A(\xi,\eta)$ defined by 
Eq. (\ref{eq-Asol}) satisfies the equation
$\partial_{\xi\eta}A=A$ which is equivalent to Eq. (\ref{eq-Axt}) for $\beta=1$.
If $\rho(+\infty)=0$ (the incident wave comes from vacuum), then
\begin{equation}\label{eq-vars}
\xi = -\frac{\omega_{pe}^2}{4}\!\!\int_{x/c-t}^{+\infty}\!\!\!\!\!\rho(\zeta)d\zeta,\;\;\;\;
\eta=\frac{x}{c}+t.
\end{equation}
The correspondence between the independent variables $(x,t)$ and $(\zeta,\eta)$,
and the region of influence of Eq. \ref{eq-Axt} for $\beta=1$
are shown in Fig. \ref{fig-0}.
The simplest bounded profile function vanishing for $x-ct>0$ is
represented by the Heaviside step function $\theta$:
\begin{align}\label{eq-rho-linear}
\rho_1(\zeta)=\theta(-\zeta). 
\end{align}

\begin{figure}
\includegraphics[width=0.8\columnwidth]{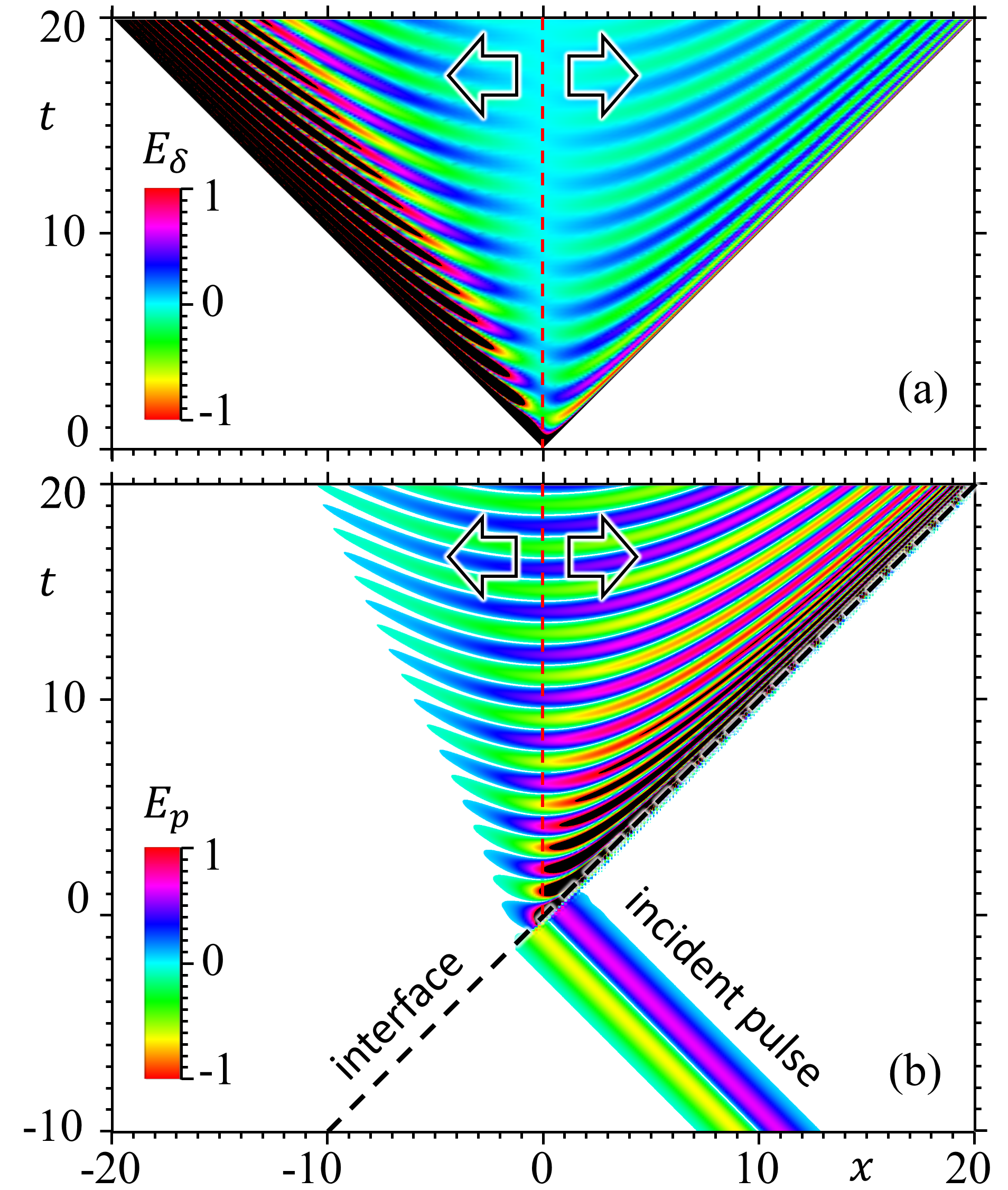}
\caption{%
The electric field strength in the $(x,t)$ plane
computed from (a) Eq. (\ref{eq-Asol-delta})
and (b) Eq. (\ref{eq-E-spec}), for $\xi,\eta$ defined by Eqs. (\ref{eq-vars}), (\ref{eq-rho-linear}),
$\alpha_\delta=1,\alpha_p=1,\omega_{pe}^2=10, \tau=1$.
Colorscale is saturated; black is for values beyond the range.
Diagonal dashed line in (b) is the plasma-vacuum interface worldline.
Vertical dashed lines correspond to zero wave group velocity of Eq.  (\ref{eq-Vph}).
Hollow arrows show the wave propagation direction.%
\label{fig-1}
}
\end{figure}
\begin{figure}
\includegraphics[width=0.8\columnwidth]{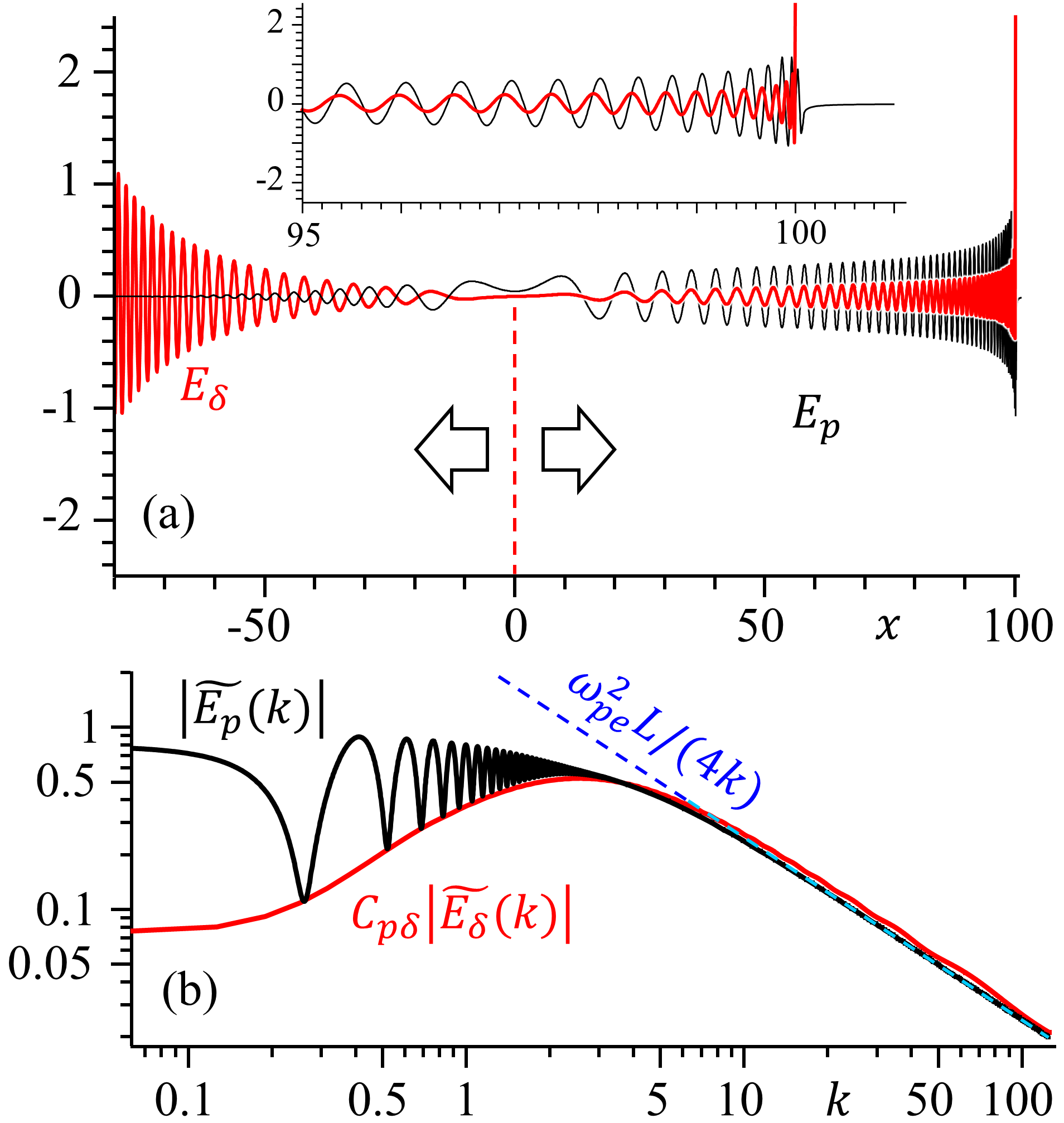}
\caption{%
The electric field strength profiles (a) and spectra (b) at $t=100$
computed from Eq. (\ref{eq-Asol-delta}), red,
and Eq. (\ref{eq-E-spec}), black, for $\alpha_\delta=1,\alpha_p=1,\omega_{pe}^2=10, \tau=1$.
The inset in (a) shows a zoomed region near $x=100$.
Vertical dashed line shows where the wave group velocity of Eq.  (\ref{eq-Vph}) is zero. 
Hollow arrows show the wave propagation direction.
Dashed line in (b) is the asymptote for large $k$.
$C_{p\delta}=\int|\widetilde{E}_p|dk/\int|\widetilde{E}_\delta|dk$.%
\label{fig-2}
}
\end{figure}

{\bf Plane wave.}
When the incident radiation is a plane wave, $A_{0w}(\xi,\eta)=\alpha_w\exp(-i\omega\eta)$,
its Fourier transform is $a_{0w}(\Omega) = \sqrt{2\pi}\;\alpha_w\delta(\Omega-\omega)$.
Then Eq. (\ref{eq-Asol}) reduces to:
%
\begin{equation}\label{eq-Asol-plane0}
A_{w}(\xi,\eta) = 
\alpha_w\exp[i\Psi(\xi,\eta)], \Psi = {\xi}/\omega - \omega\eta ,
\end{equation}
where $\xi$ and $\eta$ are given by Eq. (\ref{eq-vars}).
We see that for $x\rightarrow+\infty$, $A_{w}=A_{0w}$.
The frequency and wavenumber of the wave in Eq. (\ref{eq-Asol-plane0}),
\begin{align}\label{eq-Ow}
\Omega_w=-\partial\Psi/\partial t={\textstyle\frac{\omega_{pe}^2}{4\omega}\rho\left(\frac{x}{c}-t\right)} + \omega,
\\ \label{eq-Kw}
K_w = \partial\Psi/\partial x={\textstyle\frac{1}{c}}\left({\textstyle\frac{\omega_{pe}^2}{4\omega}\rho\left(\frac{x}{c}-t\right)} - \omega\right),
\end{align}
satisfy the dispersion relation
$\Omega_w^2=c^2 K_w^2+\omega_{pe}^2\rho(\zeta)$.
The phase and group velocity of the wave, 
$v_{ph} \!=\!{\Omega_w}/{K_w}$ and
$v_{gr} \!=\!{\partial\Omega_w}/{\partial K_w}$, are
\begin{align}\label{eq-Vph}
v_{ph}
\!=\!\frac{\omega_{pe}^2\rho\!+\!4\omega^2}{\omega_{pe}^2\rho\!-\!4\omega^2}c,\;
v_{gr}
\!=\!\frac{\omega_{pe}^2\rho\!-\!4\omega^2}{\omega_{pe}^2\rho\!+\!4\omega^2}c.
\end{align}
For $\omega>\omega_{pe}\sqrt{\rho}/2$ both velocities are negative, 
i.e. the wave propagates in the same direction as the incident wave;
if $\rho$ is bounded, the wave is transmitted with a phase shift of
$\omega_{pe}^2(4\omega)^{-1}\int_{-\infty}^{+\infty}(\rho(\zeta)-\rho_1(\zeta))d\zeta$. 
For $\omega=\omega_{pe}\sqrt{\rho}/2$ the wave becomes stationary;
here $\Omega_w$ is minimum: $\Omega_{w, {\rm min}}=\omega_{pe}\sqrt{\rho}$.
For $0<\omega<\omega_{pe}\sqrt{\rho}/2$ both velocities are positive,
i.e. the incident wave undergoes total internal reflection with frequency $\omega_r=\Omega_w> \omega_{pe}\sqrt{\rho}$,
\begin{align}\label{eq-Or}
\omega_r/\omega = 1+\omega_{pe}^2\rho/(4\omega^2) = 1+n_e/(4n_{cr})>2,
\end{align}
where $n_e\!=\! m_e\omega_{pe}^2\rho/(4\pi e^2)$ and $n_{cr}\!=\! m_e\omega^2/(4\pi e^2)\!<\! n_e/4$
are the electron density and critical density, respectively.
Quite surprisingly, the reflected wave frequency tends to infinity
when the incident frequency decreases to 0
whereas the reflection coefficient remains $R=1$.

The reflection becomes inevitable in the case of unbounded $\rho$;
e.g. for $\rho(\zeta)=-\mu\zeta\theta(-\zeta)$, it occurs at 
$t^* = t_0 + 4\omega^2/(\mu\omega_{pe}^2)$,
$x^* = t_0 c + 4(1-\ln 4)\omega^2/(\mu\omega_{pe}^2)$,
where $t_0$ is the time of the wave and interface collision.

{\bf Unit impulse.}
Another limiting case is an infinitely short pulse represented by the Dirac delta function:
$A_{0\delta}(\xi,\eta)=\alpha_\delta\delta(\eta)$ with the Fourier transform of $a_{0\delta}(\Omega) = \alpha_\delta/\sqrt{2\pi}$.
In this case the Cauchy principal value of the integral in Eq. (\ref{eq-Asol}) is \cite{GR}
\begin{equation}\label{eq-Asol-delta}
A_{\delta}(\xi,\eta) = -\alpha_\delta {J}_1(2\sqrt{-\xi\eta})\sqrt{-\xi/\eta},
\end{equation}
where ${J}_1$ is the Bessel function of the first kind.
The solution is defined only in the physically reasonable quadrant $\{\xi<0,\eta>0\}$.
It is unbounded at $\eta=0$ and
remains fixed at $\xi=0$: $A_{\delta}(\xi\rightarrow 0,\eta>0)=\alpha_\delta$.
For fixed $\xi<0$ and $\eta\rightarrow\infty$, its magnitude decreases as
$|A_{\delta}|\approx \alpha_\delta\pi^{-1/2}(-\xi)^{1/4}\eta^{-3/4}$.
If $\rho(\zeta)=0$ for $\zeta>0$, then Eq. (\ref{eq-vars}) entails
$\xi\approx (\omega_{pe}^2/4)(x/c-t)$ for $x\rightarrow ct-0$.
In this limit the solution's frequency and wavenumber increase with time (satisfying the dispersion relation):
\begin{align}\label{eq-}
\Omega_\delta\approx-\partial\left(2\sqrt{-\xi\eta}\right)/\partial t\approx -\omega_{pe}t/\sqrt{t^2-x^2/c^2},
\\
K_\delta\approx\partial\left(2\sqrt{-\xi\eta}\right)/\partial x\approx -\omega_{pe}x/(c\sqrt{t^2-x^2/c^2}).
\end{align}
For $\rho=\rho_1$ defined by Eq. (\ref{eq-rho-linear}),
Fig. \ref{fig-1}(a) and Fig. \ref{fig-2} show
the electric field strength properties computed from Eq. (\ref{eq-Asol-delta}):
the distribution in the $(x,t)$ plane, and
the profile and spectral density at $t=100$, for $\alpha_\delta=1, \omega_{pe}^2=10$.

{\bf Gaussian pulse.}
The simplest profile function, Eq. (\ref{eq-rho-linear}),
facilitates an analytic derivation of the Fourier transform of Eq. (\ref{eq-Asol})
with respect to $x/c$ in the case of a Gaussian incident pulse, $A_{0p}(\xi,\eta)=\alpha_p e^{-\eta^2/(2\tau^2)}$.
The Fourier transform of the corresponding electric field strength $E_p=-\partial A_p/\partial t$ is
\begin{align}\label{eq-E-spec}
\tilde E_p(k,t) \!=\!\alpha_p(F_k^{-K} \!+\! F_k^{K} \!-\! F_k^{k} \!+\!
i\tau k e^{-\frac{1}{2}\tau^2 k^2 - itk}),
 \\ \nonumber
F_k^\varkappa=e^{itk-2t^2/\tau^2}\zeta_k^\varkappa {w}\left[i\sqrt{2}\left(\zeta_k^\varkappa-t/\tau\right)\right],
\\ \nonumber
\zeta_k^\varkappa={i\tau}(k+\varkappa)/{4},\;
K=\sqrt{k^2+\omega_{pe}^2},
\end{align}
where ${w}(z)=e^{-z^2}{\rm erfc}(-iz)$ is the Faddeeva function \cite{FAD}.
As expected, for $t\rightarrow-\infty$, survives only the last term in the sum;
it is the Fourier transform with respect to $x/c$ of the incident Gaussian pulse electric field
$E_{0p}=-\partial A_{0p}/\partial t$.
In the opposit limit, 
\begin{equation}\label{eq-E-spec-limit}
\tilde E_p(k,t)\approx \alpha_p\frac{i\tau\omega_{pe}^2}{4k} e^{it\sqrt{k^2+\omega_{pe}^2}},\;
k\rightarrow\infty, t\gg \tau .
\end{equation}
This asymptote is the inverse of wavenumber; it is
peculiar to the Fourier transform of the Heaviside step function.
The spectral density is constant, while the phase is proportional to time and
asymptotically is proportional to wavenumber.
Thus, the energy stored in each wavenumber $k$ is constant, so the shape of the 
corresponding wavepacket in the time domain evolves only because of the phase.
Consequently, the wavepacket should contain oscillations with indefinitly increasing frequency.
For $\rho=\rho_1$ defined by Eq. (\ref{eq-rho-linear}),
Fig. \ref{fig-1}(b) and Fig. \ref{fig-2} show
the electric field strength properties computed from Eq. (\ref{eq-E-spec}):
the distribution in the $(x,t)$ plane, and
the profile and spectral density at $t=100$.
Figs. \ref{fig-1} and \ref{fig-2} compare the outcomes of the incident unit impulse and Gaussian pulse;
their respective electric field maxima near $x=t$ are 
$E_{\delta,{\rm max}}=\alpha_\delta\omega_{pe}^2/4$ for all $t$ and 
$E_{p,{\rm max}}\approx 0.12\alpha_p\tau\omega_{pe}^2$ for $t=100$
($\alpha_\delta$ and $\alpha_p$ have different dimensions by definition).
Surprisingly, the corresponding reflected waves are almost in phase (except the interval $90<x<100$)
while the transmitted waves are in antiphase.
The transmitted wave from the Gaussian pulse vanishes at $x\rightarrow -t$,
while that from the unit impulse is unbounded.

\begin{figure}
\includegraphics[width=0.8\columnwidth]{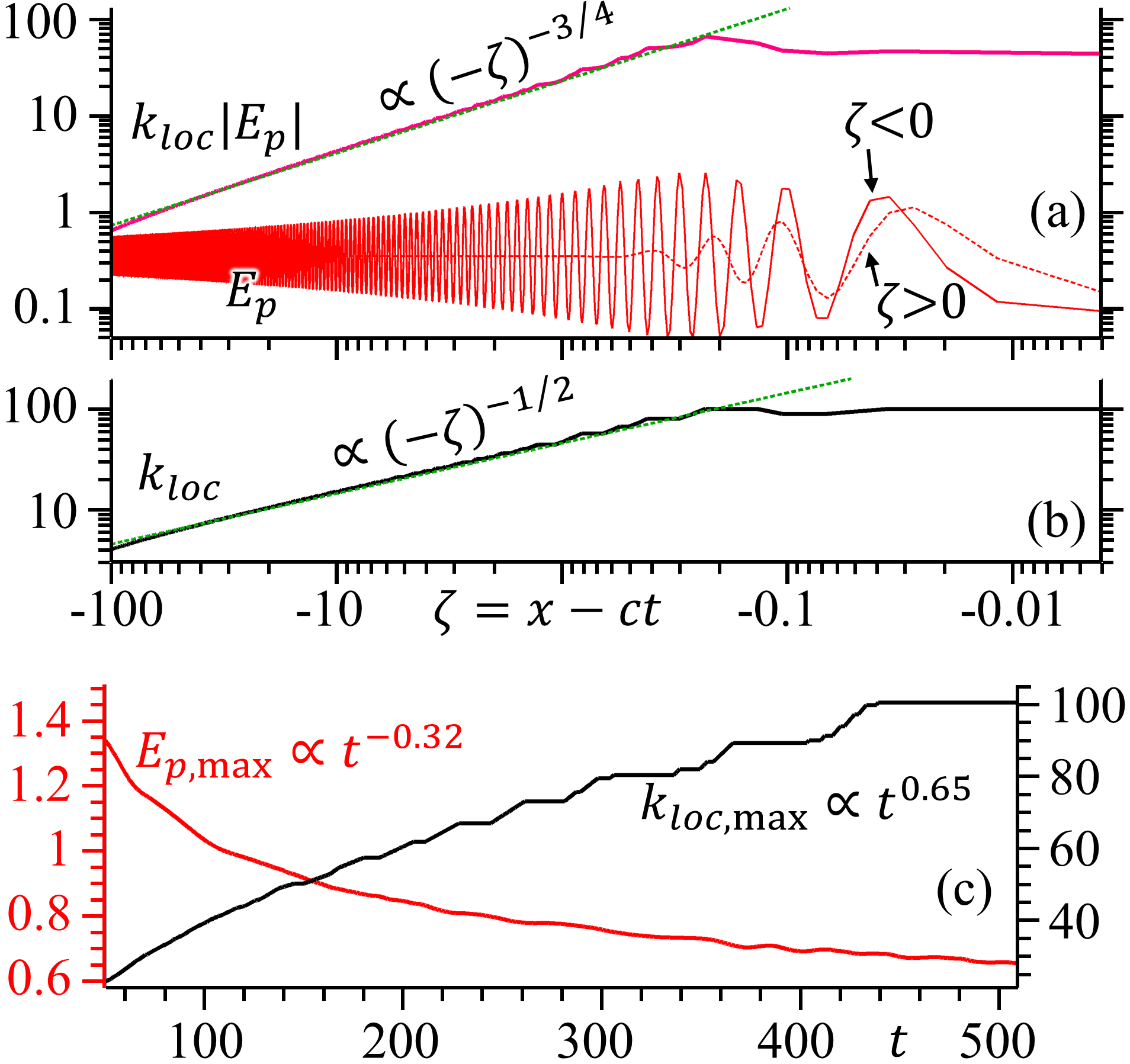}
\caption{%
Reflected wave behind the luminal plasma-vacuum interface seen in the numerical solution of Eq. (\ref{eq-Axt}).
(a) The electric field strength $E_p$ and its magnitude product $k_{loc}|E_p|$ with
(b) its local wavenumber $k_{loc}$ at $t=500$, shown for $\zeta=x-ct<0$.
In (a), $E_p$ is linearly scaled but its ordinate axis is not shown, $E_{p,{\rm max}}\approx0.67$;
the solid and dashed curves are $\{\zeta,E_p(\zeta<0)\}$ and $\{-\zeta,E_p(\zeta>0)\}$, respectively.
(c) The evolution of the maximum electric field strength (left axis) and wavenumber (right axis).%
\label{fig-3}
}
\end{figure}

{\bf Numerical solution.}
The incident Gaussian pulse has a wide spectrum with an exponentially vanishing tail,
in contrast to the non-exponential spectrum seen in Eq. (\ref{eq-E-spec}) and Fig. \ref{fig-2}(b).
Does the $k^{-1}$ dependence necessitate a discontinuous profile function like $\rho=\rho_1$?
In order to answer this question we numerically solve Eq. (\ref{eq-Axt}),
using the finite-difference time-domain (FDTD) method (e.g. see \cite{Taflove}),
for $\beta=1$, $\omega_{pe}^2=10$, $\rho(\zeta) = 0.5(1-\tanh[5\zeta])$,
and the Gaussian incident pulse with $\alpha_p=1,\tau=1$.
Fig. \ref{fig-3} confirms the asymptotic behaviour of the electric field strength near the interface
predicted above in the case of unit pulse; 
for this purpose we show the electric field $E_p$ and its local wavenumber $k_{loc}$ inside plasma behind the luminal interface in log-log scale.
We compute $k_{loc}$ as an inverse distance between adjacent local extrema
therefore it always shows the maximum frequency.
In Fig. \ref{fig-3}(a) $E_p$ extends a bit beyond $x=ct$ because of the extents of the Gaussian pulse and profile function.
Fig. \ref{fig-3}(c) shows that near the luminal interface the maximum wavenumber grows indefinitely while
the maximum electric field strenght vanish; their product increases.

\begin{figure}
\includegraphics[width=0.8\columnwidth]{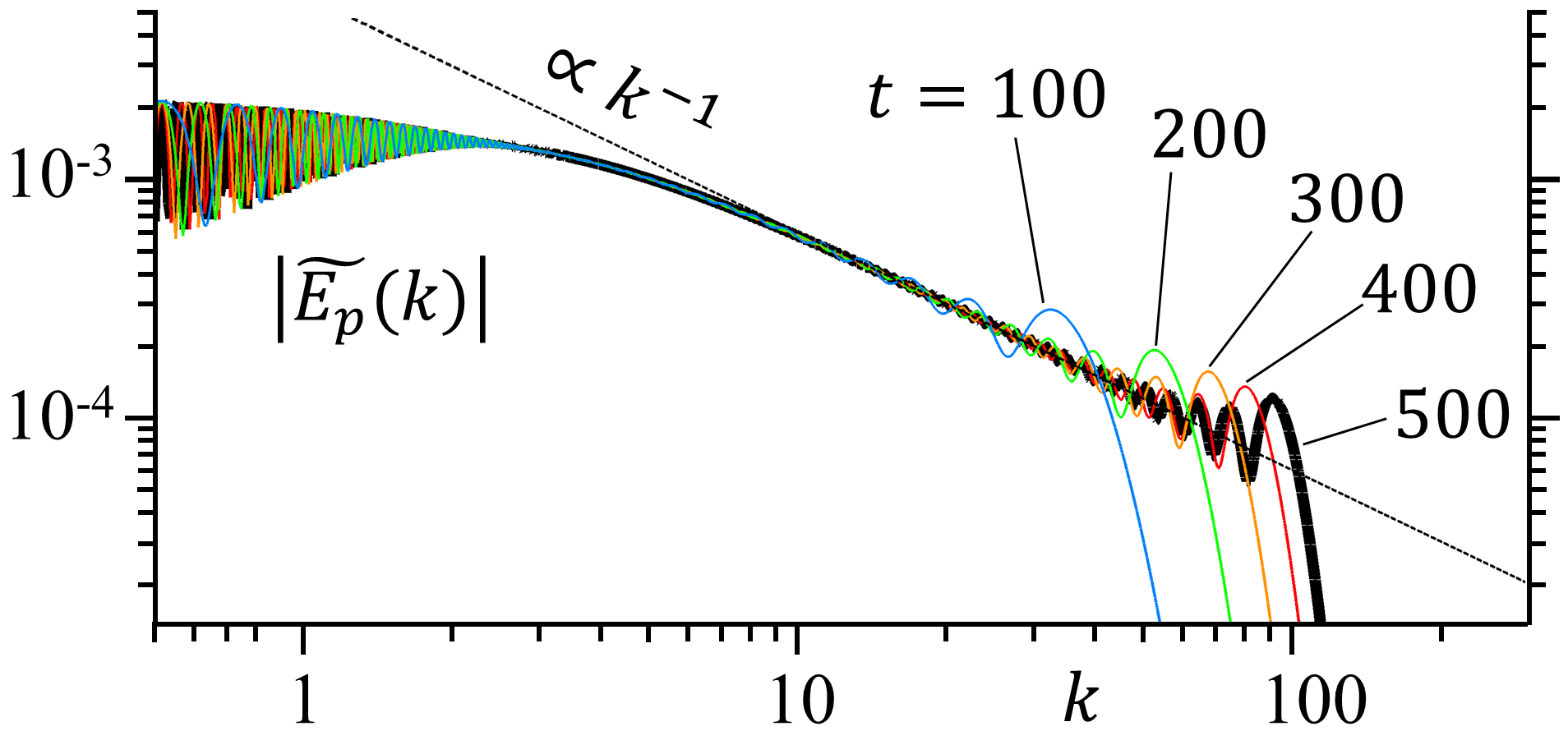}
\caption{%
The electric field strength spectra at different time moments
seen in the same numerical solution as in Fig. \ref{fig-3}.%
\label{fig-4}
}
\end{figure}

Fig. \ref{fig-4} shows the spectrum similarly to Fig. \ref{fig-2}(b),
for different moments of time.
Although the characteristic wavenumber induced by $\rho$ is $\sim 5\times 2\pi$,
we see much higher wavenumbers arranged along the $k^{-1}$ dependence.
The energy for higher wavenumbers at the next time moment
is apparently taken from the spectral maximum near the cutoff at the previous time moment.


\begin{figure}
\includegraphics{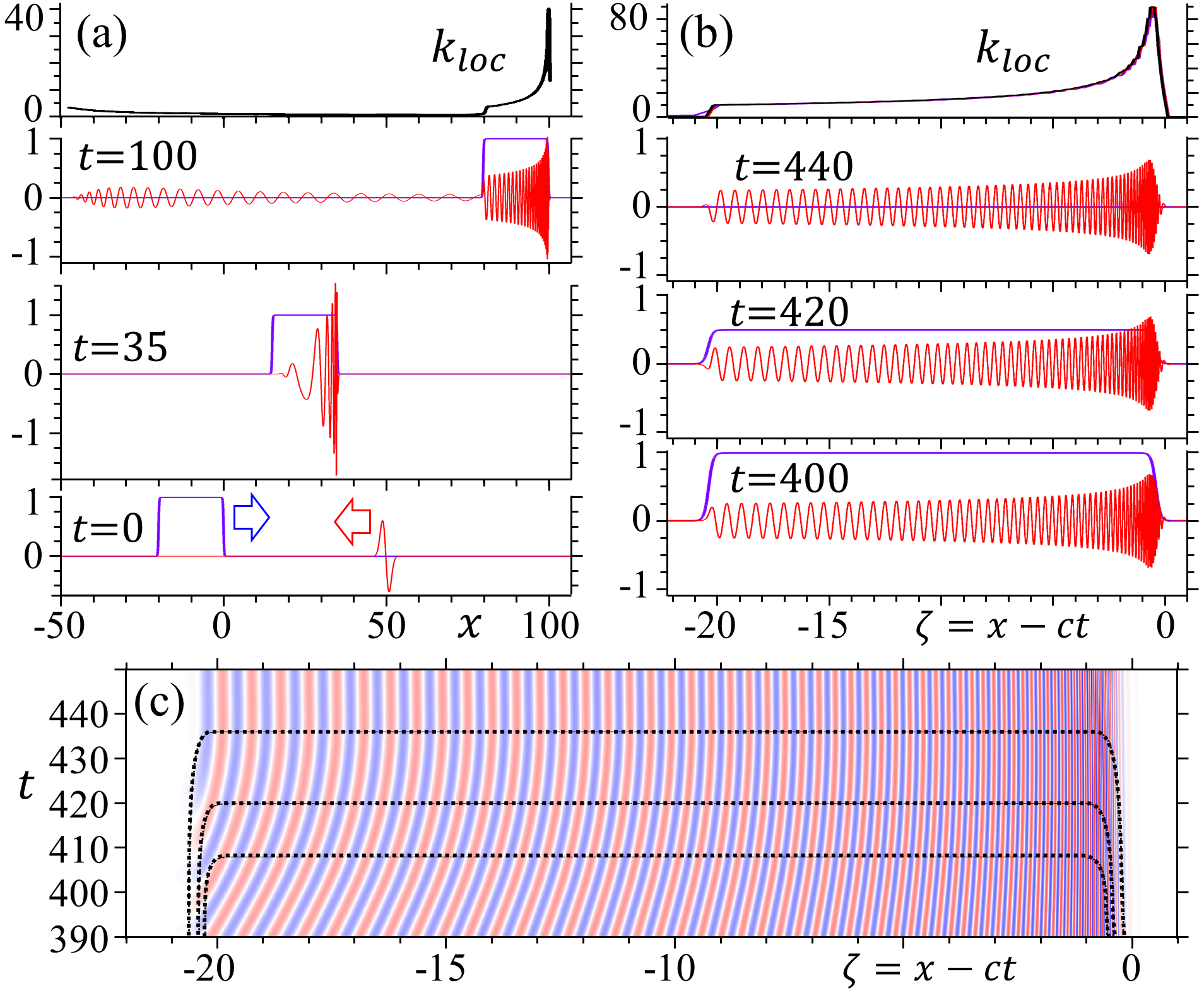}
\caption{%
The numerical solution of Eq. (\ref{eq-Axt}) for the luminal plasma slab gradually disappearing from $t=400$ till $t=440$.
(a) The beginning of the Gaussian pulse and plasma slab interaction,
and the local wavenumber of the latest instance of the electric field (top).
(b) The electric field in the disappearing plasma slab,
and the almost coinciding profiles of the electric field local wavenumber
(blue,red,black for $t=400,420,440$).
(c) The electric field strength and the profile function isocurves $\rho=0.8, 0.5, 0.1$ in the $(x-ct,t)$ plane.%
\label{fig-5}
}
\end{figure}

{\bf Disappearing plasma slab.}
A phase object (in particular, the luminal plasma slab) can disappear almost instantaneously
if an agent causing the local modification of medium properties ceases its action.
For example, consider an intense focusing searchlight inducing a change of the refractive 
index of ambient gas at the moving focus;
when it is turned off, the refractive index returns to its normal value.
This rise the question: if the luminal plasma slab disappears
(i.e. the modulations of its refractive index vanish),
what will happen to the high frequency radiation stored in the slab?
We performed numerical solution of Eq. (\ref{eq-Axt}) with the luminal plasma slab of thickness 20.
Its profile function $\rho(\zeta)=0.5(\tanh[5(\zeta+20)]-\tanh[5\zeta])$
starts to linearly decrease at $t_X=400$ and vanishes at $t_X+\Delta t_X=440$.
The reflection of the Gaussian pulse in such the slab is seen Fig. \ref{fig-5}.
The reflected wave stays inside the slab, near the interface its magnitude decreases and frequency increases in time
with the rate shown in Fig. \ref{fig-3}(c).
The local wavenumber is almost preserved during the luminal plasma slab disappearance, Fig. \ref{fig-5}(b).
The electromagnetic field strength distribution in the $(x-ct,t)$ plane, Fig. \ref{fig-5}(c),
reveals a change of the equal-phase curves.
Inside the luminal plasma slab, they are inclined;
the inclination angle tangent is $1/(\beta_{ph}-1)$
with the wave phase velocity $\beta_{ph}$ defined by Eq. (\ref{eq-Vph}).
When the profile function $\rho$ vanishes,
the equal-phase curves becomes straight vertical as it should be in vacuum in the $(x-ct,t)$ plane. 

{\bf Conclusion.}
When a refractive index modulation moves at the speed of light in vacuum,
an incident electromagnetic wave, depending on its frequency, either is totally transmitted with a modified phase,
or forms a standing wave,
or is totally reflected with an upshifted frequency.
A sufficiently low-frequency incident plane wave is reflected whithout losses
and is compressed by the factor of $1+n_e/(4n_{cr})>2$.
All three parts -- transmitted, standing and reftected -- present in a long wavepacket
produced by a short incident electromagnetic pulse behind the luminal plasma-vacuum interface.
The reflected part has the asymptotic energy spectral density inversely proportional to the frequency squared, while its local wavenumber near the interface grows indefinitely.
If the luminal plasma slab containing the reflected wave disappears,
the high frequency radiation is released as a short chirped wavepacket.

The authors thank F. Pegoraro, A. S. Pirozhkov, M. Kando, P. V. Sasorov, and P. Valenta for fruitful discussions. 


\end{document}